\documentclass{article}

\usepackage{color}

\def\b{\begin{equation}}
\def\e{\end{equation}}

\def\ba{\begin{eqnarray}}
\def\ea{\end{eqnarray}}
\usepackage{latexsym}
\usepackage{amsfonts}

\title{\bf On the autonomisation of differential equations I:
Sundman transformations for Painlev\'e equations I-IV}

\author{\\   {\bf Pilar R. Gordoa, Alexandru Iosif and Andrew Pickering} \\ \\
             \'Area de Matem\'atica Aplicada, ESCET \\
             Universidad Rey Juan Carlos \\
             C/~Tulip\'an s/n, 28933 M\'ostoles \\
             Madrid, Spain}

\date{9 September 2026}

\begin{document}

\maketitle

\thispagestyle{empty}

\begin{tabbing}

\noindent \smallskip
{\em Short title:} \= On the autonomisation of differential equations \\


\noindent \bigskip
{\em MSC2020 classification scheme numbers:} \> {}\hskip 46mm

37K35, 34M55, 33E17, 34M15

% 33E17 Painlev\'e-type functions
% 37K35 Lie-B\"acklund and other transformations for infinite dimensional 
%       Hamiltonian and Lagrangian systems
% 34M55 Painlev\´e and other special ordinary differential equations in the 
%       complex domain; classification, hierarchies
% 34M15 Algebraic aspects (differential-algebraic, hypertranscendence, 
%       group-theoretical) of ordinary differential equations in the complex 
%       domain

\end{tabbing}

\begin{abstract}

We consider the application of Sundman transformations to the first four 
Painlev\'e equations $P_I$-$P_{IV}$, with $P_V$ and $P_{VI}$ being left for
future work. Since the Painlev\'e equations are nonautonomous, this requires 
first of all the construction of an equivalent autonomous system, a process 
which we discuss in detail, and which we characterize as being inverse to 
the use of Lie symmetries to obtain a reduction of order of an autonomous 
system. Secondly, we apply a Sundman transformation to this autonomous system,
and identify the resulting equation with a corresponding nonautonomous equation. 
We give an explicit mapping between solutions of this nonautonomous equation 
and the original Painlev\'e equation. Interesting special cases of the 
nonautonomous equation found by Sundman transformation include, for $P_{II}$, 
an equation with coefficients expressed via Airy functions, and for $P_{IV}$, 
an equation with coefficients expressed via Weber-Hermite functions. We also 
discuss an alternative autonomisation process. Finally, a variety of possible 
directions for future work are identified, along with interesting open problems. 

\end{abstract}

\vfill

{\bf Corresponding author:} A.~Pickering. email: andrew.pickering@urjc.es

\newpage

\setcounter{section}{0} \setcounter{equation}{0}

\section{Introduction}

The main aim of this paper is to explore the application of Sundman transformations
to the Painlev\'e equations. Here we will consider the first four Painlev\'e
equations, $P_I$-$P_{IV}$, leaving $P_V$ and $P_{VI}$ to a later paper. A secondary
but very closely related aim is the derivation of autonomous equations, and in
particular of autonomous polynomial systems, equivalent to the Painlev\'e equations.

Sundman transformations were first introduced in \cite{Sund13}, see also 
\cite{DMS94}, and have since been widely employed in the study of solutions of
differential equations arising in a wide variety of contexts, see, e.g.,
\cite{KS14}---\cite{GPPT25}. More recent generalisations of Sundman transformations, 
and their application to the NRT nonlinear Schr\"odinger equation, can be found in
\cite{GPPT26}. However, the Painlev\'e equations are nonautonomous differential
equations, and so Sundman transformations cannot be applied to them directly. 
Instead, as observed in \cite{E07}, for the particular case of the first Painlev\'e
equation $P_I$, a first step of passing to an equivalent autonomous equation needs 
to be taken. In \cite{E07}, this was achieved by using a hodograph transformation. 
Thus, for $P_I$,
\b
y_{zz}=6y^2+z,
\e
where we use subscripts to denote derivatives, the authors of \cite{E07} first used
a hodograph transformation to obtain a corresponding third order differential
equation. They then applied a Sundman transformation to this third order equation,
and thus obtained a mapping from solutions of an Emden-Fowler equation to $P_I$.

In this paper we put this approach in a more general context, and provide an
interpretation of this first step of obtaining an equivalent autonomous system
as a process inverse to the use of Lie symmetries to reduce the order of an 
autonomous equation (and passing to a nonautonomous equation). This is discussed 
in Section 2. Also in Section 2, with an eye on future work, we consider the
derivation of equivalent polynomial autonomous systems. In Section 3, we apply
the techniques of Section 2 to $P_I$-$P_{IV}$. For $P_I$ a special case of our
results is the third order autonomous equation to which Sundman transformations
were applied in \cite{E07}. In Section 4 we consider the application of Sundman
transformations not only to this equation, but also to corresponding equations
obtained in Section 3 for $P_{II}$-$P_{IV}$. It is very interesting that certain
equations well-known for their relation to Painlev\'e equations also make an
appearance in this context. For example, for $P_{II}$, the Airy equation appears,
with the end result being a mapping from a nonautonomous Duffing equation to $P_{II}$.
In Section 5 we consider an alternative means of deriving autonomous equations
equivalent to the Painlev\'e equations, as well as the further application of
Sundman transformations. In the final Section 6 we draw our conclusions, as well
as discussing possible future directions of research and questions arising from the
results derived here.

\setcounter{equation}{0}

\section{On the autonomisation of systems of equations}

\subsection{From nonautonomous to autonomous equations}

\noindent {\bf Theorem 2.1}
Given any system of $n$ nonautonomous first order ordinary differential equations
\b
y_{i,z}=f_i(z,y_1,y_2,\ldots,y_n),\qquad{}i=1,2,\ldots,n,
\label{naut}
\e
where the subscript $z$ denotes the derivative with respect to the independent
variable $z$, there exists a transformation to a corresponding equivalent system 
of $n+1$ autonomous ordinary differential equations
\b
v_{i,x}=g_i(v_0,v_1,v_2,\ldots,v_n),\qquad{}i=0,1,\ldots,n,
\label{aut}
\e
where the subscript $x$ denotes the derivative with respect to the new independent variable $x$, and where we are free to choose the evolution of $v_0$ (i.e., we are
free to choose $g_0(v_0,v_1,v_2,\ldots,v_n)\not\equiv 0$).

\vskip 2mm

\noindent {\em Proof:} We make the change of variables 
\b
z=v_0(x),\qquad{} v_i(x)=y_i(v_0(x)),\ \  i=1,2,\ldots,n,
\e 
which then gives
\b
v_{i,x}=y_{i,z}v_{0,x}=f_i(v_0,v_1,v_2,\ldots,v_n)v_{0,x},\qquad{} i=1,2,\ldots,n.
\e
If, in addition, we choose the evolution of $v_0$ to be given by
$v_{0,x}=g_0(v_0,v_1,v_2,\ldots,v_n)$ for some function $g_0\not\equiv 0$, we obtain 
the system (\ref{aut}) with $g_i=f_ig_0$, $i=1,2,\ldots,n$.

\noindent $\Box$

\vskip 2mm

As should be expected, we have the following result:

\vskip 2mm

\noindent {\bf Theorem 2.2} All autonomous systems (\ref{aut}) obtained from
(\ref{naut}), as described in Theorem 2.1, are equivalent.

\vskip 2mm

\noindent {\em Proof:} Beginning with the system (\ref{naut}), we consider the
alternative transformation
\b
z=w_0(t),\qquad{} w_i(t)=y_i(w_0(t)),\ \  i=1,2,\ldots,n.
\e
Choosing $w_{0,t}=h_0(w_0,w_1,w_2,\ldots,w_n)$, we thus obtain the autonomous system
\ba
w_{0,t} & = & h_0(w_0,w_1,w_2,\ldots,w_n), \label{syst1} \\
w_{i,t} & = & h_i(w_0,w_1,w_2,\ldots,w_n) =
f_i(w_0,w_1,w_2,\ldots,w_n)h_0(w_0,w_1,w_2,\ldots,w_n), \ \  i=1,2,\ldots,n.
\label{syst2}
\ea
This system is equivalent to the system obtained in Theorem 2.1, i.e.,
\ba
v_{0,x} & = & g_0(v_0,v_1,v_2,\ldots,v_n), \label{sysx1} \\
v_{i,x} & = & g_i(v_0,v_1,v_2,\ldots,v_n) = 
f_i(v_0,v_1,v_2,\ldots,v_n)g_0(v_0,v_1,v_2,\ldots,v_n), \ \  i=1,2,\ldots,n.
\label{sysx2}
\ea
In order to see this, we make the (implicit) change of independent variable
$w_0(t)=v_0(x)$, which implies $w_i(t)=v_i(x)$, $i=1,2,\ldots,n$.
We also see that $w_{0,t}=v_{0,x}\frac{dx}{dt}$, so
$\frac{dx}{dt}=h_0(w_0(t),\ldots,w_n(t))/g_0(v_0(x),\ldots,v_n(x))$. Thus
$w_{0,t}=v_{0,x}[h_0(w_0(t),\ldots,w_n(t))/g_0(v_0(x),\ldots,v_n(x))]$ and
$w_{i,t}=v_{i,x}[h_0(w_0(t),\ldots,w_n(t))/g_0(v_0(x),\ldots,v_n(x))]$, 
$i=1,2,\ldots,n$. It then follows that the change of variable $w_0(t)=v_0(x)$
transforms (\ref{sysx1}), (\ref{sysx2}) into (\ref{syst1}), (\ref{syst2}).

\noindent $\Box$

\vskip 2mm

\noindent {\bf Remark 2.3} The above process of autonomisation of a given
nonautonomous system, with a consequent increase in the order of the system 
by one, is the opposite to that whereby we may use the translation invariance 
of a given autonomous system to reduce its order by one by transforming to a
nonautonomous system.

\vskip 2mm

\noindent {\bf Remark 2.4} The above-mentioned process of reduction of
order is as follows. In the autonomous system (\ref{sysx1}), (\ref{sysx2}) 
we make the change of variables $z=v_0(x)$ and $y_i(z)=v_i(x(z))$, 
$i=1,2,\ldots,n$. This gives the $n$-component system
$y_{i,z}=v_{i,x}/v_{0,x}=v_{i,x}/g_0(v_0,v_1,v_2,\ldots,v_n)=
f_i(v_0,v_1,v_2,\ldots,v_n)=f_i(z,y_1,y_2,\ldots,y_n)$, $i=1,2,\ldots,n$.

\subsection{Polynomial autonomous systems}

We now discuss the derivation of equivalent polynomial autonomous systems, 
since in this case we can, moreover, apply differential elimination theory 
\cite{CF06}, a topic to which we hope to return in future papers.

\vskip 2mm

\noindent {\bf Corollary 2.5} If in the system (\ref{naut}) each $f_i$ is
a rational function of its arguments, then we may obtain a corresponding
autonomous system (\ref{aut}) where each $g_i$, $i=0,1,2,\ldots,n$, is 
polynomial in its arguments.

\vskip 2mm

\noindent {\em Proof:} It is enough to take $g_0$ to be the product of all
of the denominators of the rational functions $f_i$. We may alternatively 
take $g_0$ to be the product of the least powers of suitable factors such 
that all $f_ig_0$ are polynomial. In the latter case, the system (\ref{aut}) 
is uniquely determined up to an overall constant factor. In particular, since 
any such factor just corresponds to a rescaling of $x$, in the case where all 
$f_i$ are already polynomial, we may take $g_0=1$.

\noindent $\Box$

\vskip 2mm

\noindent {\bf Corollary 2.6} In the special case where all $f_i$ are polynomial
in $1/v_k$, for some common subset $\{v_k\}$ of the variables $v_i$, $0\leq i\leq n$,
with coefficients rational in the remaining variables $v_j$, an alternative means of
obtaining a system (\ref{aut}) with right-hand-sides polynomial in all variables is 
to take $g_0$ to be independent of the variables $v_k$ in question but such that the
right-hand-sides of the system (\ref{aut}) are polynomial in the remaining variables
$v_j$, and then map each of the variables $v_k$ to $u_k=1/v_k$ (so
$v_{k,x}=-u_{k,x}/u_k^2$) whilst renaming the
remaining $v_j$ as $u_j$.

\vskip 2mm

Considering once again the system (\ref{naut}) of $n$ nonautonomous equations,
in the case where each $f_i$ is a rational function of its arguments, an 
alternative approach to the construction of an equivalent polynomial system is 
as given in Theorem 2.7 below. The price to be paid, however, is that when in 
the original system (\ref{naut}) at least one $f_i$ is non-polynomial, the 
order of the resulting equivalent polynomial system is greater than $n+1$.

\vskip 2mm

\noindent {\bf Theorem 2.7} In the system (\ref{naut}) we assume each
$f_i(z,y_1,y_2,\ldots,y_n)=P_i(z,y_1,y_2,\ldots,y_n)/Q_i(z,y_1,y_2,\ldots,y_n)$,
where all $P_i$ and $Q_i$ are polynomial functions of their arguments. Denoting
derivations with respect to $z$ by $\partial$, the system (\ref{naut}) is 
equivalent to the autonomous polynomial system
\ba
\partial y_{i} & = & P_i(x,y_1,y_2,\ldots,y_n)w_i,\qquad{}i=1,2,\ldots,n, 
\label{np1} \\
\partial w_{i} & = & -w_i^2q_i(x,y_1,y_2,\ldots,y_n),\quad{}\ i=1,2,\ldots,n, 
\label{np2} \\
\partial x & = & 1, 
\label{np3}
\ea
subject to the initial conditions at some point of analyticity of solutions $z_0$,
\b
\begin{array}{rl}
\frac{1}{w_i}\,  \vline\,_{_{_{z=z_0}}} = Q_i\,  \vline\,_{_{_{z=z_0}}},\quad i=1,2,\ldots,n,
\qquad{}   {\rm and} \qquad{} x\, \vline\,_{_{_{z=z_0}}}=z_0,
\label{cnds}
\end{array}
\e
where each $q_i=\partial Q_i$, and where in the case that for a set of $r$ integers
$k\in\{1,2,\ldots,n\}$ we have each $Q_k=1$ (i.e., $f_k$ is in fact polynomial), 
the corresponding $r$ equations $\partial w_{k}= -w_k^2q_k$ are not to be included 
in (\ref{np2}).

\vskip 2mm

\noindent {\em Proof:} We set $x=z$, so that $\partial x=1$, and also $w_i=1/Q_i$,
$i=1,2,\ldots,n$, so that $\partial w_i=-(\partial Q_i)/Q_i^2=-w_i^2 q_i$. If some
$Q_k=1$, then $w_k=1$ and so $\partial w_k=0$ and also $q_k=\partial Q_k=0$, making
trivial the equation $\partial w_{k}= -w_k^2q_k$.
The $n+1-r$ initial conditions (\ref{cnds}) guarantee 
the equivalence of (\ref{naut}) and (\ref{np1})---(\ref{np3}), i.e., that not 
only is (\ref{np1})---(\ref{np3}) obtained from (\ref{naut}), but also that
precisely the system (\ref{naut}) is obtained from (\ref{np1})---(\ref{np3}).

\noindent $\Box$

\vskip 2mm

\noindent {\bf Remark 2.8} In the case where all $Q_i=1$, $i=1,2,\ldots,n$, the
resulting system (\ref{np1}), (\ref{np3}) (we have $r=n$ and so (\ref{np2}) is
excluded) is equivalent to the autonomous system of order $n+1$ obtained using 
Theorem 2.1 in the case $g_0=1$. In general, the system (\ref{np1})---(\ref{np3}) 
is of order $2n+1-r$ for some integer $r\in\{1,2,\ldots,n\}$.

\vskip 2mm

\noindent {\bf Remark 2.9} With a view to future work, we set $y_i=x_i$,
$i=1,2,\ldots,n$, $w_i=x_{n+i}$, $i=1,2,\ldots,n$, and $x=x_{2n+1}$, and
so rewrite the system of $2n+1-r$ equations (\ref{np1})---(\ref{np3}) as
\ba
\partial x_{i} & = & P_i(x_{2n+1},x_1,x_2,\ldots,x_n)x_{n+i},\qquad{}
i=1,2,\ldots,n, \label{np1a} \\
\partial x_{n+i} & = & -x_{n+i}^2q_i(x_{2n+1},x_1,x_2,\ldots,x_n),\quad{}\ 
i=1,2,\ldots,n, \label{np2a} \\
\partial x_{2n+1} & = & 1, \label{np3a}
\ea
and the $n+1-r$ conditions (\ref{cnds}) as
\b
\begin{array}{rl}
	\frac{1}{x_{n+i}}\,  \vline\,_{_{_{z=z_0}}} = Q_i\,  \vline\,_{_{_{z=z_0}}},\quad i=1,2,\ldots,n,
	\qquad{}   {\rm and} \qquad{} x_{2n+1}\, \vline\,_{_{_{z=z_0}}}=z_0.
	\label{cndsa}
\end{array}
\e

\vskip 2mm

\noindent {\bf Remark 2.10} Setting $x=z$, and augmenting a system with the
equation $\partial x=1$ (or $\partial x_{2n+1}=1$ above), clearly transforms any
nonautonomous system into an autonomous system. Whilst this may seem in some 
sense artificial, we recall that it is precisely this approach that has been
successfully used in the study of ``fully noncommutative'' equations \cite{RR10}. 
Our results here correspond to its adoption in the fully commutative case.

\setcounter{equation}{0}

\section{Examples: the Painlev\'e equations}

The six Painlev\'e equations have the form $y_{zz}=F(z,y,y_z)$, for some 
rational function $F$ of $z$, $y$ and $y_z$. Setting $y=y_1$ and $y_z=y_2$,
in the usual way, we can rewrite these equations as a system
\ba
y_{1,z} & = & y_2, \label{pnaut1} \\
y_{2,z} & = & F(z,y_1,y_2). \label{pnaut2}
\ea
Applying Theorem 2.1 and Corollary 2.5 to this system, we obtain 
equivalent third order autonomous systems
\ba
v_{0,x} & = & g_0(v_0,v_1,v_2), \label{paut1} \\
v_{1,x} & = & g_0(v_0,v_1,v_2) v_2, \label{paut2} \\
v_{2,x} & = & g_0(v_0,v_1,v_2) F(v_0,v_1,v_2), \label{paut3}
\ea
where we may choose $g_0$ in such a way that equations
(\ref{paut1})---(\ref{paut3}) have polynomial right-hand-sides.

\vskip 2mm

\noindent {\bf Remark 3.1}
In the reverse process, i.e., that of reducing by one the order of a 
given autonomous system by transforming to a nonautonomous system, as
described in Remark 2.4, we consider the autonomous system
(\ref{paut1})---(\ref{paut3}) and make the change of variables $z=v_0(x)$,
$y_1(z)=v_1(x(z))$, $y_2(z)=v_2(x(z))$, which yields (\ref{pnaut1}), 
(\ref{pnaut2}):
$y_{1,z}=v_{1,x}/v_{0,x}=v_{1,x}/g_0=v_2=y_2$ and
$y_{2,z}=v_{2,x}/v_{0,x}=v_{2,x}/g_0=F(v_0,v_1,v_2)=F(z,y_1,y_2)$.

\vskip 2mm

In the remainder of this section we apply the above ideas to the first four
Painlev\'e equations $P_{I}$-$P_{IV}$. For reasons of simplicity, we will
consider in the main examples where $g_0(v_0,v_1,v_2)=v_1^p$, for some nonzero
integer $p$. Particular attention will be paid to the resulting third order 
differential equation satisfied by $v_0$, as it is to this equation, for the
particular choice $p=-1$, that we will later apply Sundman transformations. We
make this choice in order to facilitate a comparison with the results given
in \cite{E07} for $P_I$. In addition in this section we will give results
for the case $g_0(v_0,v_1,v_2)=1$, as well as for the application of Theorem
2.7 to $P_{III}$ and $P_{IV}$.

\subsection{The first Painlev\'e equation}

The first Painlev\'e equation, $P_I$, has the form
\b
y_{zz}=6y^2+z,
\e
and so in the corresponding system (\ref{pnaut1}), (\ref{pnaut2}) we have
$F(z,y_1,y_2)=6y_1^2+z$. For the choice $g_0=v_1^p$, $p\neq0$, we obtain 
using Theorem 2.1 the equivalent autonomous system
\ba
v_{0,x} & = & v_1^p, \label{p14a} \\
v_{1,x} & = & v_1^p v_2, \\
v_{2,x} & = & v_1^p(6v_1^2+v_0). \label{p14c}
\ea
Elimination between these equations yields the following third order equation for 
$v_0$:
\b
\frac{v_{0,xxx}}{v_{0,x}}-\frac{2p-1}{p}\frac{v_{0,xx}^2}{v_{0,x}^2}=
p(v_{0,x})^\frac{2p-1}{p}\left(v_0+6(v_{0,x})^\frac{2}{p}\right).
\label{p14}
\e
For the choice $p=1$, for example, the system (\ref{p14a})---(\ref{p14c}) is 
polynomial. We note also that for $p=2$ equation (\ref{p14}) is expressed 
using the Schwarzian derivative:
\b
\frac{v_{0,xxx}}{v_{0,x}}-\frac{3}{2}\frac{v_{0,xx}^2}{v_{0,x}^2}=
2(v_{0,x})^\frac{3}{2}(v_0+6v_{0,x}).
\label{p1sch}
\e
For this choice $p=2$ the system (\ref{p14a})---(\ref{p14c}) is again polynomial.
The question of differential equations involving the Schwarzian derivative related
to Painlev\'e equations was first explored by Weiss \cite{W84}, for the second
Painlev\'e equation. In this section we give autonomous examples of such equations, 
for $P_I$ (as above), $P_{II}$ and $P_{IV}$.

The choice $p=-1$ corresponds to the particular autonomization of $P_I$ used in
\cite{E07}: the system (\ref{p14a})---(\ref{p14c}) is non-polynomial,
\ba
v_{0,x} & = & \frac{1}{v_1}, \label{p13a} \\
v_{1,x} & = & \frac{v_2}{v_1}, \\
v_{2,x} & = & 6v_1+\frac{v_0}{v_1}, \label{p13c}
\ea
and the corresponding third order equation (\ref{p14}) now reads
\b
v_{0,xxx}=3\frac{v_{0,xx}^2}{v_{0,x}}-6v_{0,x}^2-v_0v_{0,x}^4,
\label{p13}
\e
which is precisely the autonomous equation related to $P_I$ given in \cite{E07}.
In Section 4 we will consider, as in \cite{E07}, the application of Sundman transformations to equation (\ref{p13}). Similarly, with the aim of adopting a
consistent approach in our study of $P_{II}$-$P_{IV}$, we will also consider in
Section 4 the application of Sundman transformations to the third order equations
in $v_0$ obtained for these equations using $g_0=1/v_1$.

Finally, we remark that choosing $g_0=1$ yields the polynomial system
\ba
v_{0,x} & = & 1, \\
v_{1,x} & = & v_2, \\
v_{2,x} & = & 6v_1^2+v_0,
\ea
which is the same as would also be obtained using Theorem 2.7.

\subsection{The second Painlev\'e equation}

For the second Painlev\'e equation, $P_{II}$, i.e.,
\b
y_{zz}=2y^3+zy+\alpha,
\e
where $\alpha$ is an arbitrary constant, in the corresponding system 
(\ref{pnaut1}), (\ref{pnaut2}) we have $F(z,y_1,y_2)=2y_1^3+zy_1+\alpha$.
For this equation, the choice $g_0=v_1^p$, $p\neq0$ leads us to the 
autonomous system
\ba
v_{0,x} & = & v_1^p, \label{p2a} \\
v_{1,x} & = & v_1^p v_2, \\
v_{2,x} & = & v_1^p(2v_1^3+v_0v_1+\alpha),\label{p2c}
\ea
which in turn yields the following third order equation for $v_0$:
\b
v_{0,xxx}=\left(\frac{2p-1}{p}\right)\frac{v_{0,xx}^2}{v_{0,x}}+\alpha p (v_{0,x})^\frac{3p-1}{p}+2p(v_{0,x})^\frac{3p+2}{p}+pv_0v_{0,x}^3.
\e
As for $P_I$, we see that for $p=1$ and $p=2$, for example, the system
(\ref{p2a})---(\ref{p2c}) is polynomial and, once again, for the latter
choice, the third order equation in $v_0$ involves the Schwarzian derivative:
\b
\frac{v_{0,xxx}}{v_{0,x}}-\frac{3}{2}\frac{v_{0,xx}^2}{v_{0,x}^2}=
2\alpha (v_{0,x})^\frac{3}{2}+4v_{0,x}^3+2v_0v_{0,x}^2.
\e

For the choice $p=-1$ we obtain the non-polynomial system
\ba
v_{0,x} & = & \frac{1}{v_1},  \\
v_{1,x} & = & \frac{v_2}{v_1}, \\
v_{2,x} & = & 2v_1^2+v_0+\frac{\alpha}{v_1},
\ea
and the corresponding third order equation for $v_0$,
\b
\frac{v_{0,xxx}}{v_{0,x}}-3\frac{v_{0,xx}^2}{v_{0,x}^2}=-2-v_0v_{0,x}^2-\alpha v_{0,x}^3,\label{p23}
\e
which is the equation that we will use later in our indirect application, via
autonomisation, of Sundman transformations to $P_{II}$.

Let us also write down the autonomous system obtained by taking $g_0=1$:
\ba
v_{0,x} & = & 1, \\
v_{1,x} & = & v_2, \\
v_{2,x} & = & 2v_1^3+v_0v_1+\alpha,
\ea
again as would also be obtained using Theorem 2.7.

\subsection{The third Painlev\'e equation}

The third Painlev\'e equation, $P_{III}$, has the form
\b
y_{zz}=\frac{y_z^2}{y}-\frac{y_z}{z}+\frac{1}{z}(\alpha y^2+\beta)+\gamma y^3
+\frac{\delta}{y},
\e
where $\alpha$, $\beta$, $\gamma$ and $\delta$ are arbitrary constants,
and so, in the corresponding system (\ref{pnaut1}), (\ref{pnaut2}), 
we have $\displaystyle F(z,y_1,y_2)=\frac{y_2^2}{y_1}-\frac{y_2}{z}
+\frac{1}{z}(\alpha y_1^2+\beta)+\gamma y_1^3+\frac{\delta}{y_1}$.

We see that, in order to obtain a polynomial system, we may take $g_0=v_1v_0$ 
to obtain
\ba
v_{0,x} & = & v_1v_0, \\
v_{1,x} & = & v_1 v_0 v_2, \\
v_{2,x} & = & v_2^2 v_0-v_2 v_1
+(\alpha v_1^3+\beta v_1)+\gamma v_1^4 v_0+\delta v_0,
\ea
and the corresponding equation for $v_0$,
\b
\frac{v_{0,xxx}}{v_{0,x}}-2\frac{v_{0,xx}^2}{v_{0,x}^2}+\frac{v_{0,xx}}{v_0}=
\alpha \frac{v_{0,x}^3}{v_0^2}+\gamma \frac{v_{0,x}^4}{v_0^2}+\beta v_{0,x}+\delta v_0^2.
\e
Alternatively, noting that $F$ is polynomial in $1/z$, and so in $1/v_0$,
we see from Corollary 2.6 that we may take $g_0=v_1$ to obtain
\ba
v_{0,x} & = & v_1, \\
v_{1,x} & = & v_1 v_2, \\
v_{2,x} & = & v_2^2 -\frac{v_2 v_1}{v_0}
+\frac{1}{v_0}(\alpha v_1^3+\beta v_1)+\gamma v_1^4+\delta,
\ea
and then make the further change of variables $v_0=1/u_0$, $v_1=u_1$, $v_2=u_2$,
in order to obtain the polynomial system
\ba
u_{0,x} & = & -u_1 u_0^2, \\
u_{1,x} & = & u_1 u_2, \\
u_{2,x} & = & u_2^2 -u_2 u_1 u_0
+u_0(\alpha u_1^3+\beta u_1)+\gamma u_1^4+\delta,
\ea
with the equation satisfied by the variable $u_0$ being
\b
\frac{u_{0,xxx}}{u_{0,x}}-2\frac{u_{0,xx}^2}{u_{0,x}^2}+\frac{u_{0,xx}}{u_0}=
-\alpha \frac{u_{0,x}^3}{u_0^5}+\gamma \frac{u_{0,x}^4}{u_0^8}-\beta \frac{u_{0,x}}{u_0}+\delta.
\e

More generally, the choice $g_0=v_1^p$ yields the system
\ba
v_{0,x} & = & v_1^p, \\
v_{1,x} & = & v_1^p v_2, \\
v_{2,x} & = & v_1^{p-1}\left(v_2^2 -\frac{v_2 v_1}{v_0}
+\frac{1}{v_0}(\alpha v_1^3+\beta v_1)+\gamma v_1^4+\delta\right),
\ea
and the corresponding equation for $v_0$
\b
\frac{v_{0,xxx}}{v_{0,x}}-2\frac{v_{0,xx}^2}{v_{0,x}^2}+\frac{v_{0,xx}}{v_0}=
\alpha p \frac{(v_{0,x})^{(2p+1)/p}}{v_0}+\gamma p (v_{0,x})^{(2p+2)/p}+\beta p \frac{(v_{0,x})^{(2p-1)/p}}{v_0}+\delta p (v_{0,x})^{(2p-2)/p}.
\e
We note that, in contrast to $P_I$ and $P_{II}$, and also, as we shall soon
see, $P_{IV}$, for $P_{III}$ it is not possible for this choice of $g_0$ to 
choose $p$ in such a way that the equation in $v_0$ involves the Schwarzian 
derivative.

Taking $p=-1$ gives the system
\ba
v_{0,x} & = & \frac{1}{v_1}, \\
v_{1,x} & = & \frac{v_2}{v_1}, \\
v_{2,x} & = & \frac{v_2^2}{v_1^2} -\frac{v_2}{v_0v_1}
+\frac{1}{v_0v_1}(\alpha v_1^2+\beta )+\gamma v_1^2+\frac{\delta}{v_1^2},
\ea
along with the corresponding third order equation for $v_0$,
\b
\frac{v_{0,xxx}}{v_{0,x}}-2\frac{v_{0,xx}^2}{v_{0,x}^2}+\frac{v_{0,xx}}{v_0}=
-\alpha  \frac{v_{0,x}}{v_0}-\gamma -\beta  \frac{v_{0,x}^3}{v_0}-\delta  v_{0,x}^{4}.
\label{p33}
\e
It is to this last equation that, in the next section, we will apply Sundman
transformations.

We note that for $P_{III}$ the choice $g_0=1$ does not lead to a polynomial 
autonomous system. Instead, we obtain
\ba
v_{0,x} & = & 1, \\
v_{1,x} & = & v_2, \\
v_{2,x} & = & \frac{v_2^2}{v_1}-\frac{v_2}{v_0}
+\frac{1}{v_0}(\alpha v_1^2+\beta)+\gamma v_1^3+\frac{\delta}{v_1}.
\ea
This system may be compared to the fourth order system obtained using Theorem 
2.7, given in Section 3.5, which also involves a variable with derivative 
defined to be $1$.

\subsection{The fourth Painlev\'e equation}

The fourth Painlev\'e equation, $P_{IV}$, reads
\b
y_{zz}=\frac{1}{2}\frac{y_z^2}{y}+\frac{3}{2}y^3+4zy^2+2(z^2-\alpha)y
+\frac{\beta}{y},
\label{p4}
\e
where $\alpha$ and $\beta$ are arbitrary constants,
and so, in the corresponding system (\ref{pnaut1}), (\ref{pnaut2}), we have
$\displaystyle F(z,y_1,y_2)=\frac{1}{2}\frac{y_2^2}{y_1}+\frac{3}{2}y_1^3+4zy_1^2+2(z^2-\alpha)y_1+\frac{\beta}{y_1}$.
We find that the choice $g_0=v_1^p$ leads to the autonomous system
\ba
v_{0,x} & = & v_1^p, \label{p4a}\\
v_{1,x} & = & v_1^pv_2, \\
v_{2,x} & = & v_1^p\left(\frac{1}{2}\frac{v_2^2}{v_1}+\frac{3}{2}v_1^3+4v_0v_1^2+2(v_0^2-\alpha)v_1+\frac{\beta}{v_1}\right),\label{p4c}
\ea
from which it follows that $v_0$ satisfies the third order equation
\b
\frac{v_{0,xxx}}{v_{0,x}}-\frac{4p-1}{2p}\frac{v_{0,xx}^2}{v_{0,x}^2}=
\frac{3}{2}p(v_{0,x})^{(2p+2)/p}+\beta p (v_{0,x})^{(2p-2)/p}
-2\alpha p v_{0,x}^2+2pv_0^2v_{0,x}^2+4pv_0(v_{0,x})^{(2p+1)/p}.
\e
We note that for $p=1$, this last equation is expressed in terms of the 
Schwarzian derivative,
\b
\frac{v_{0,xxx}}{v_{0,x}}-\frac{3}{2}\frac{v_{0,xx}^2}{v_{0,x}^2}=
\frac{3}{2}v_{0,x}^4+\beta
-2\alpha  v_{0,x}^2+2v_0^2v_{0,x}^2+4v_0v_{0,x}^3,
\e
with the corresponding system (\ref{p4a})---(\ref{p4c}) being polynomial.

For the choice $p=-1$ we obtain from (\ref{p4a})---(\ref{p4c}) the autonomous system
\ba
v_{0,x} & = & \frac{1}{v_1}, \\
v_{1,x} & = & \frac{v_2}{v_1}, \\
v_{2,x} & = & \frac{1}{2}\frac{v_2^2}{v_1^2}+\frac{3}{2}v_1^2+4v_0v_1+2(v_0^2-\alpha)
+\frac{\beta}{v_1^2},
\ea
with the corresponding equation for $v_0$, to which we will apply Sundman 
transformations, being given by
\b
\frac{v_{0,xxx}}{v_{0,x}}-\frac{5}{2}\frac{v_{0,xx}^2}{v_{0,x}^2}=
-\frac{3}{2}-\beta v_{0,x}^4
+2\alpha  v_{0,x}^2-2v_0^2v_{0,x}^2-4v_0v_{0,x}.\label{p43}
\e

We may also take $g_0=1$ to obtain the non-polynomial system
\ba
v_{0,x} & = & 1, \\
v_{1,x} & = & v_2, \\
v_{2,x} & = & \frac{1}{2}\frac{v_2^2}{v_1}+\frac{3}{2}v_1^3+4v_0v_1^2+2(v_0^2-\alpha)v_1
+\frac{\beta}{v_1}.
\ea
As for $P_{III}$, we remark that the above system may be compared 
to the fourth order system obtained using Theorem 2.7, which we give in Section 3.5,
which also involves a variable with derivative defined to be $1$.

\subsection{Fourth order autonomous equations equivalent to $\mathbf{P_{III}}$ and $\mathbf{P_{IV}}$}

We now consider the results of applying Theorem 2.7 to $P_{III}$ and $P_{IV}$.

In the case of $P_{III}$ we take $P_1=y_2$, $Q_1=1$ and $w_1=1/Q_1=1$, and 
$P_2=y_2^2z-y_1y_2+y_1(\alpha y_1^2+\beta)+\gamma y_1^4 z+\delta z$, $Q_2=y_1z$
and $w_2=1/Q_2=1/y_1z$. The resulting fourth order system in $x_1=y_1$, $x_2=y_2$,
$x_4=w_2$ and $x_5=z$ is
\ba
\partial x_1 & = & x_2, \\
\partial x_2 & = & \Big(x_2^2x_5-x_1x_2+x_1(\alpha x_1^2+\beta)+\gamma x_1^4 x_5 
+\delta x_5 \Big)x_4, \\
\partial x_4 & = & -x_4^2(x_2x_5+x_1), \\
\partial x_5 & = & 1,
\ea
subject to the constraints that at some point of analyticity of solutions $z_0$,
\b
\begin{array}{rl}\displaystyle
	\frac{1}{x_4}\,  \vline\,_{_{_{z=z_0}}} = x_1x_5\,  \vline\,_{_{_{z=z_0}}}
	\qquad{}   {\rm and} \qquad{} x\, \vline\,_{_{_{z=z_0}}}=z_0.
\end{array}
\e

For $P_{IV}$ we may take $P_1=y_2$, $Q_1=1$ and $w_1=1/Q_1=1$, and 
$P_2=\frac{1}{2}y_2^2+\frac{3}{2}y_1^4+4y_1^3z+2(z^2-\alpha)y_1^2+\beta$, $Q_2=y_1$
and $w_2=1/Q_2=1/y_1$. The resulting fourth order system in $x_1=y_1$, $x_2=y_2$,
$x_4=w_2$ and $x_5=z$ is
\ba
\partial x_1 & = & x_2, \\
\partial x_2 & = & \Big(\frac{1}{2}x_2^2+\frac{3}{2}x_1^4+4x_1^3x_5+2(x_5^2-\alpha)x_1^2+\beta \Big)x_4, \\
\partial x_4 & = & -x_4^2x_2, \\
\partial x_5 & = & 1,
\ea
subject to the constraints that at some point of analyticity of solutions $z_0$,
\b
\begin{array}{rl}\displaystyle
	\frac{1}{x_4}\,  \vline\,_{_{_{z=z_0}}} = x_1\,  \vline\,_{_{_{z=z_0}}}
	\qquad{}   {\rm and} \qquad{} x\, \vline\,_{_{_{z=z_0}}}=z_0.
\end{array}
\e
We will return to the study of these polynomial systems, which are equivalent to $P_{III}$ and 
$P_{IV}$, in later papers.

\setcounter{equation}{0}

\section{The Painlev\'e equations and Sundman transformations}

In the previous section we derived, for each of the Painlev\'e equations 
$P_I$-$P_{IV}$, a variety of equivalent third order autonomous equations
in the variable $v_0(x)$ (which is in fact the independent variable of the
corresponding Painlev\'e equation). We will now consider the equations in
$v_0$ obtained by making the choice $g_0=1/v_1$, and will apply Sundman
transformations to these equations. For the particular case of $P_I$, this
is as done in \cite{E07}, in order to establish a relationship between $P_I$
and an Emden-Fowler equation. For reasons of completeness, we recover this
result here. The results given here for $P_{II}$-$P_{IV}$ are, however, 
all new.

After applying a Sundman transformation, which yields a new equation in a
function $X(t)$, we construct a new equivalent second order nonautonomous 
equation in a function $W(X)$, related to the equation in $X(t)$ in the
same way that our equation in $v_0(x)$ is related to its Painlev\'e equation
in $y(z)$. That is, we proceed in the following way:

\vskip 2mm

\noindent {\bf Sundman transformations}
Applying the Sundman transformation
\b
v_0(x)=F(X),\qquad  dx=G(X)dt, \qquad {\rm where} \qquad GF_X\neq0,
\label{sund}
\e
to a third order autonomous equation in $v_0(x)$ yields a new third order
autonomous equation in $X(t)$. This can be written as a system in $X$, $Y$
and $Z$ by setting $X_t=Y$ and $Y_t=Z$, with the evolution of $Z$ being
determined by the third order equation, in the usual way. From this system
a nonautonomous equation in $Y(X)$ can then be derived. However, in order
to obtain a nonautonomous equation related to the third order equation in 
$X(t)$ in the same way as our third order equation in $v_0$ is related to
its Painlev\'e equation, we make the further change of variables $Y=1/W$.
These steps are best understood by considering the examples given below.
Furthermore, we see that solutions of the equation in $W(X)$ are mapped to
solutions of the Painlev\'e equation via the mapping
\b
y(z)=\frac{WG}{F_X},\qquad{} z=F(X),
\label{smap}
\e
since $\displaystyle y=y_1=v_1=\frac{1}{v_{0,x}}=\frac{1}{F_{X}}\frac{dx}{dX}$ and
$\displaystyle \frac{dx}{dX}=\frac{dx}{dt}\frac{dt}{dX}=G\frac{1}{Y}=GW$. Also,
since $z=v_0(x)$, we have $z=F(X)$.
We note that the explicit inverse mapping (\ref{smap}) is not to be found in \cite{E07},
but rather an integral version thereof.

\vskip 2mm

The sequence of changes of variable from a Painlev\'e equation in $y(z)$ to the 
equation in $W(X)$ is as follows:

\vskip 5mm

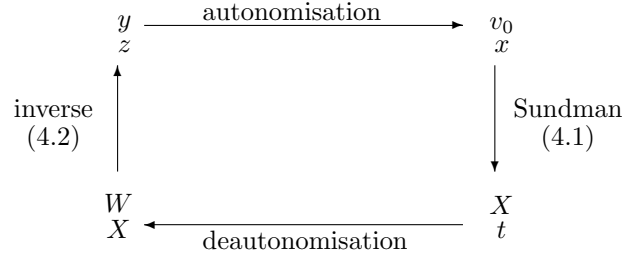
\begin{center}

\begin{picture}(200,80)
\put(72,77){autonomisation}
\put(50,75){\vector(1,0){120}}
\put(170,0){\vector(-1,0){120}}
\put(72,-9){deautonomisation}
\put(1,30){\shortstack{inverse \\ (\ref{smap})}}
\put(182,60){\vector(0,-1){40}}
\put(40,20){\vector(0,1){40}}
\put(189,30){\shortstack{Sundman \\ (\ref{sund})}}
\put(40,65){\shortstack{$y$ \\ $z$}}
\put(35,-5){\shortstack{$W$ \\ $X$}}
\put(180,65){\shortstack{$v_0$ \\ $x$}}
\put(180,-5){\shortstack{$X$ \\ $t$}}
\end{picture}

\end{center}

\vskip 5mm

\vskip 2mm

\noindent {\bf Remark 4.1} As noted in \cite{E07}, a more general Sundman 
transformation, $v_0(x)=F(X,t)$, $dx=G(X,t)dt$, with 
$G\frac{\partial F}{\partial _X}\neq0$, may also be considered. We refer 
also to \cite{gu1}. A further generalisation is considered in \cite{gu2},
although leaving the independent variable invariant. We leave the 
consideration of such generalisations to future papers.

\vskip 2mm

We now turn to the application of the Sundman transformation (\ref{sund}). In
each case, we seek simplifications of the resulting equations in $W(X)$, just 
as for the $P_I$ case considered in \cite{E07} a restriction to an Emden-Fowler 
equation was obtained. Once again, further exploration of this part of the 
process will be left to later papers.

Use will be made of the following:

\vskip 2mm

\noindent {\bf Lemma 4.2} The general solution of the Schwarzian equation
\b
\left(\frac{F_{XX}}{F_X}\right)_X-\frac{1}{2}\left(\frac{F_{XX}}{F_X}\right)^2
=\frac{1}{2}g(F)F_X^2
\label{Fsch}
\e
is given by
\b
X+c=\int\frac{dF}{\psi^2(F)},
\label{XcF}
\e
where $c$ is an arbitrary constant and $\psi(F)$ is the general solution of the 
second order linear differential equation
\b
\frac{d^2\psi}{dF^2}=\frac{1}{4}g(F)\psi.
\label{psieqn}
\e

\vskip 2mm

\noindent {\em Proof:} We set $F_X=U(F)$. Then $\displaystyle F_{XX}=U\frac{dU}{dF}$,
$\displaystyle F_{XX}/F_X=\frac{dU}{dF}$ and 
$\displaystyle (F_{XX}/F_X)_X=U\frac{d^2U}{dF^2}$. We thus obtain
\b
\frac{d}{dF}\left(\frac{1}{U}\frac{dU}{dF}\right)+\frac{1}{2}
\left(\frac{1}{U}\frac{dU}{dF}\right)^2=\frac{1}{2}g(F).
\e
The further change of variable $U(F)=\psi^2(F)$ then gives (\ref{psieqn}).
Thus, given the general solution of (\ref{psieqn}), the general solution
of (\ref{Fsch}) can be obtained from the separable equation 
$F_X=\psi^2(F)$, which then gives (\ref{XcF}).

\noindent $\Box$

\vskip 2mm

\noindent {\bf Remark 4.3} The result of Lemma 4.2 is a special case of
the solution of equation 3.5.4-1(7) given in \cite{PZ02}.

\vskip 2mm

\subsection{The first Painlev\'e equation}

We consider equation (\ref{p13})
\b
v_{0,xxx}=3\frac{v_{0,xx}^2}{v_{0,x}}-6v_{0,x}^2-v_0v_{0,x}^4,
\label{p1v0}
\e
and perform a Sundman transformation of the form (\ref{sund}), i.e.,
\b
v_0(x)=F(X),\qquad  dx=G(X)dt, \qquad {\rm where} \qquad GF_X\neq0,
\e
We find that the new independent variable $X(t)$ satisfies the third-order equation
\b
X_tX_{ttt}-3X_{tt}^2+A(X)X_t^2X_{tt}+B(X)X_t^5+C(X)X_t^4+D(X)X_t^3=0,
\label{p1X}
\e
where
\ba
A(X) &=& 2\frac{G_X}{G}-3\frac{F_{XX}}{F_X},\\
B(X) &=& \frac{FF_X^3}{G},\\
C(X) &=& \frac{F_{XXX}}{F_X}+3\frac{G_XF_{XX}}{GF_X}-3\frac{F_{XX}^2}{F_X^2}-\frac{G_{XX}}{G},\\
D(X) &=& 6GF_X,
\ea
and where, since we must have $GF_X\neq0$, both $B(X)$ and $D(X)$ are  different from
zero. 

We now proceed as described at the beginning of this section, and write the above
equation as the system
\ba
X_t&=&Y,\label{p1s1}\\
Y_t&=&Z,\\
Z_t&=&3\frac{Z^2}{Y}-A(X)YZ-B(X)Y^4-C(X)Y^3-D(X)Y^2,\label{p1s3}
\ea
from which it follows that the equation satisfied by $Y=Y(X)$ is
\b
YY_{XX}=2Y_X^2-A(X)YY_X-B(X)Y^3-C(X)Y^2-D(X)Y.
\e
In order to obtain an equation related to (\ref{p1X}) in the same way as $P_I$
is related to (\ref{p1v0}), we make the additional change of variables $Y=1/W$,
which yields
\b
W_{XX}+A(X)W_X-B(X)-C(X)W-D(X)W^2=0.
\label{p1W}
\e

Let us remark, for this particular example, that a system for (\ref{p1X})
corresponding to (\ref{p1W}) in the same way that $P_I$ corresponds to the
system giving rise to (\ref{p1v0}), i.e., to the choice $g_0=1/v_1$, is
obtained from (\ref{p1s1})---(\ref{p1s3}) by setting $Y=1/W$ and $Z=-V/W^3$, 
which gives
\ba
X_t&=&\frac{1}{W},\\
W_t&=&\frac{V}{W},\\
V_t&=&\frac{1}{W}\left(-A(X)V+B(X)+C(X)W+D(X)W^2\right).
\ea
However, the derivation of this last system is unnecessary, given the aims
of the current paper, and it is enough to write down the system
(\ref{p1s1})---(\ref{p1s3}), obtain the equation in $Y(X)$ and then set $Y=1/W$.
This is how we will proceed in subsequent examples.

We have thus recovered the result given in \cite{E07} that autonomising $P_I$ 
(with $g_0=1/v_1$) and applying the Sundman transformation (\ref{sund}),
and then deautonomising in an equivalent way, leads to a mapping between
solutions of (\ref{p1W}) and solutions of $P_I$. This mapping is as given
in equation (\ref{smap}).

In order to relate specific equations to $P_I$, we may fix the coefficients
of (\ref{p1W}) by making choices of $F$ and $G$. We give here two illustrative
examples. In the first case, we choose to set $A(X)=C(X)=0$, which then requires
\b
F=-\frac{b}{X+c}+d,\qquad{} G=\frac{a}{(X+c)^3},
\e
where $a$, $b$, $c$ and $d$ are arbitrary constants, $ab\neq0$, for which choice 
(\ref{p1W}) becomes an Emden-Fowler equation, as observed in \cite{E07}.
Alternatively, we may choose to set $A(X)=0$ and $D(X)=6$, which requires
\b
F=\frac{X+c}{b},\qquad{} G=b,
\e
where $b$ and $c$ are arbitrary constants, $b\neq0$, and for which choice (\ref{p1W})
becomes
\b
W_{XX}=6W^2+\frac{X+c}{b^5}.
\label{p1Wb}
\e
In this second case we just obtain the combined scaling and shift transformation 
$y(z)=b^2W(X)$, $\displaystyle z=\frac{X+c}{b}$, which maps (\ref{p1Wb}) to $P_I$ 
in standard form. As such mappings are not of great interest, for the remainder
of our examples $P_{II}$-$P_{IV}$ we will seek simplifications of the equation in 
$W(X)$ by setting coefficients to be zero.

\subsection{The second Painlev\'e equation}

We now consider equation (\ref{p23})
\b
\frac{v_{0,xxx}}{v_{0,x}}-3\frac{v_{0,xx}^2}{v_{0,x}^2}=-2-v_0v_{0,x}^2-\alpha v_{0,x}^3.
\e
The Sundman transformation (\ref{sund}) yields the following equation for $X(t)$:
\b
X_tX_{ttt}-3X_{tt}^2+A(X)X_t^2X_{tt}+B(X)X_t^5+C(X)X_t^4+D(X)X_t^2=0,
\e
where
\ba
A(X) &=& 2\frac{G_X}{G}-3\frac{F_{XX}}{F_X},\\
B(X) &=&\alpha  \frac{F_X^3}{G},\\
C(X) &=& \frac{F_{XXX}}{F_X}+3\frac{G_XF_{XX}}{GF_X}-3\frac{F_{XX}^2}{F_X^2}-\frac{G_{XX}}{G}+FF_X^2,\\
D(X) &=& 2G^2,
\ea
and where we must have that, unless $\alpha=0$, both $B(X)$ and $D(X)$ are
different from zero. We now write the above equation as the system
\ba
X_t&=&Y,\\
Y_t&=&Z,\\
Z_t&=&3\frac{Z^2}{Y}-A(X)YZ-B(X)Y^4-C(X)Y^3-D(X)Y,
\ea
and we find that the equation satisfied by $Y=Y(X)$ is
\b
YY_{XX}=2Y_X^2-A(X)YY_X-B(X)Y^3-C(X)Y^2-D(X).
\e
Making the change of variable $Y(X)=1/W(X)$, we obtain the following 
equation for $W$:
\b
W_{XX}+A(X)W_X-B(X)-C(X)W-D(X)W^3=0.
\label{p2W}
\e
The Sundman transformation (\ref{smap}) thus gives a mapping from
solutions of (\ref{p2W}) to solutions of $P_{II}$.

Let us now consider the case where we impose the conditions that $A(X)=C(X)=0$.
Eliminating $G$, we find that $F(X)$ must satisfy the equation
\b
\left(\frac{F_{XX}}{F_X}\right)_X-\frac{1}{2}\left(\frac{F_{XX}}{F_X}\right)^2
=2FF_X^2,
\label{p2Fsch}
\e
which is precisely the case $g(F)=4F$ of equation (\ref{Fsch}). It then follows
from Lemma 4.2 that the solution of (\ref{p2Fsch}) is given by
\b
X+c=\int\frac{dF}{\psi^2(F)},
\label{p2XcF}
\e
where $c$ is an arbitrary constant and $\psi(F)$ is the general solution of the
linear equation
\b
\frac{d^2\psi}{dF^2}=F\psi.
\label{p2psieqn}
\e
We then find
\b
G=d\psi^3(F)
\e
for some arbitrary constant $d$, where we must have $d\psi(F)\neq0$. The Sundman
transformation (\ref{sund}) thus yields a mapping from (\ref{p2W}), i.e.,
\b
W_{XX}=\frac{\alpha}{d}\psi^3(F)+2d^2\psi^6(F)W^3,
\label{p2Wa}
\e
to $P_{II}$. We note that in this last equation four arbitrary constants appear:
$c$, $d$, and two arbitrary constants $a$ and $b$ which appear via the general 
solution of the Airy equation (\ref{p2psieqn}). It is interesting that the Airy 
equation occurs here, as this equation is well-known to play a fundamental role 
in the study of solutions of $P_{II}$. We also note that the equation obtained 
here with a role analogous to that of the Emden-Fowler equation obtained in the 
$P_{I}$ case is the nonautonomous Duffing equation (\ref{p2Wa}), or, more generally, (\ref{p2W}); for studies of such equations see, e.g., \cite{LR15}.

\subsection{The third Painlev\'e equation}

We now consider equation (\ref{p33}),
\b
\frac{v_{0,xxx}}{v_{0,x}}-2\frac{v_{0,xx}^2}{v_{0,x}^2}+\frac{v_{0,xx}}{v_0}=
-\alpha  \frac{v_{0,x}}{v_0}-\gamma -\beta  \frac{v_{0,x}^3}{v_0}-\delta  v_{0,x}^{4}.
\e
The Sundman transformation (\ref{sund}) yields the following equation for $X(t)$:
\b
X_tX_{ttt}-2X_{tt}^2+A(X)X_t^2X_{tt}+B(X)X_t^6+C(X)X_t^5+D(X)X_t^4+E(X)X_t^3+H(X)X_t^2=0,
\e
where
\ba
A(X) &=& \frac{F_X}{F}-\frac{F_{XX}}{F_X},\\
B(X) &=&\delta  \frac{F_X^4}{G^2},\\
C(X) &=&\beta \frac{F_X^3}{FG} ,\\
D(X) &=& \frac{F_{XXX}}{F_X}+\frac{G_XF_{XX}}{GF_X}-2\frac{F_{XX}^2}{F_X^2}+\frac{F_{XX}}{F}-\frac{G_{XX}}{G}+\frac{G_X^2}{G^2}-\frac{F_XG_X}{FG},\\
E(X)&=&\alpha \frac{F_XG}{F},\\
H(X)&=&\gamma G^2,
\ea
and where we must have that, unless we have a zero value for some parameter(s),
$B(X), C(X), E(X)$ and $H(X)$ are all different from zero. We now write the 
equation for $X(t)$ as the system
\ba
X_t&=&Y,\\
Y_t&=&Z,\\
Z_t&=&2\frac{Z^2}{Y}-A(X)YZ-B(X)Y^5-C(X)Y^4-D(X)Y^3-E(X)Y^2-H(X)Y,
\ea
from which we find that the equation satisfied by $Y=Y(X)$ is
\b
YY_{XX}=Y_X^2-A(X)YY_X-B(X)Y^4-C(X)Y^3-D(X)Y^2-E(X)Y-H(X),
\e
with the corresponding equation for $W(X)=1/Y(X)$ being
\b
W_{XX}=\frac{W_X^2}{W}-A(X)W_X+B(X)\frac{1}{W}+C(X)+D(X)W+E(X)W^2+H(X)W^3=0.
\e

We now consider the case where we impose that $A(X)=D(X)=0$. This then requires that
\b
\frac{F_{XX}}{F_X}=\frac{F_{X}}{F},\qquad{} \frac{G_{XX}}{G_X}=\frac{G_{X}}{G},
\e
and so
\b
F=ae^{bX},\qquad{}G=ce^{dX},
\e
where $a$, $b$, $c$ and $d$ are four arbitrary constants, $abc\neq0$. We thus obtain a
mapping to solutions of $P_{III}$ from solutions of the equation
\b
W_{XX}=\frac{W_X^2}{W}+B(X)\frac{1}{W}+C(X)+E(X)W^2+H(X)W^3,
\label{p3e1}
\e
where
\ba
B(X) &=&\delta  \frac{a^4b^4}{c^2}e^{2(2b-d)X},\label{p3e2}\\
C(X) &=&\beta \frac{a^2b^3}{c}e^{(2b-d)X},\\
E(X)&=&\alpha bce^{dX},\\
H(X)&=&\gamma c^2e^{2dX}. \label{p3e5}
\ea
For the choice $a=b=c=d=1$ we obtain the equation,
\b
W_{XX}=\frac{W_X^2}{W}+\delta e^{2X}\frac{1}{W}+\beta e^X+\alpha e^XW^2+\gamma e^{2X}W^3,
\e
which is of course a well-known alternative form of the third Painlev\'e equation,
see, e.g., \cite{GLS02}. Equation (\ref{p3e1}) with coefficients (\ref{p3e2})---(\ref{p3e5})
then provides further such alternative forms of $P_{III}$.

\subsection{The fourth Painlev\'e equation}

We now consider equation (\ref{p43}),
\b
\frac{v_{0,xxx}}{v_{0,x}}-\frac{5}{2}\frac{v_{0,xx}^2}{v_{0,x}^2}=
-\frac{3}{2}-\beta v_{0,x}^4
+2\alpha  v_{0,x}^2-2v_0^2v_{0,x}^2-4v_0v_{0,x}.
\e
The Sundman transformation (\ref{sund}) yields the following equation for $X(t)$:
\b
X_tX_{ttt}-\frac{5}{2}X_{tt}^2+A(X)X_t^2X_{tt}+B(X)X_t^6+C(X)X_t^4+D(X)X_t^3+E(X)X_t^2=0,
\e
where
\ba
A(X) &=& \frac{G_X}{G}-2\frac{F_{XX}}{F_X},\\
B(X) &=&\beta  \frac{F_X^4}{G^2},\\
C(X) &=& \frac{F_{XXX}}{F_X}+2\frac{G_XF_{XX}}{GF_X}-\frac{5}{2}\frac{F_{XX}^2}{F_X^2}-\frac{G_{XX}}{G}+\frac{1}{2}\frac{G_X^2}{G^2}+2F^2F_X^2-2\alpha F_X^2,\\
D(X)&=&4GFF_X,\\
E(X)&=&\frac{3}{2} G^2,
\ea
and where we must have that, unless $\beta=0$, $B(X), D(X)$ and $E(X)$ are all 
different from zero. We now proceed as for our previous examples and write down 
the equivalent system
\ba
X_t&=&Y,\\
Y_t&=&Z,\\
Z_t&=&\frac{5}{2}\frac{Z^2}{Y}-A(X)YZ-B(X)Y^5-C(X)Y^3-D(X)Y^2-E(X)Y.
\ea
The equation satisfied by $Y=Y(X)$ is found to be
\b
YY_{XX}=\frac{3}{2}Y_X^2-A(X)YY_X-B(X)Y^4-C(X)Y^2-D(X)Y-E(X),
\e
and the equation satisfied by $W(X)=1/Y(X)$ is then
\b
W_{XX}=\frac{1}{2}\frac{W_X^2}{W}-A(X)W_X+B(X)\frac{1}{W}+C(X)W+D(X)W^2+E(X)W^3.
\label{p4W}
\e

Let us now consider the case where we impose $A(X)=C(X)=0$. Eliminating $G$, we 
find that $F$ must satisfy the Schwarzian equation
\b
\left(\frac{F_{XX}}{F_X}\right)_X-\frac{1}{2}\left(\frac{F_{XX}}{F_X}\right)^2
=2(F^2-\alpha)F_X^2,
\label{p4Fsch}
\e
which is of the form (\ref{Fsch}) with $g(F)=4(F^2-\alpha)$. From Lemma 4.2 we 
then obtain that the general solution of (\ref{p4Fsch}) is given by
\b
X+c=\int\frac{dF}{\psi^2(F)},
\label{p4XcF}
\e
where $c$ is an arbitrary constant and $\psi(F)$ is the general solution of the
linear equation
\b
\frac{d^2\psi}{dF^2}=(F^2-\alpha)\psi.
\label{p4psieqn}
\e
We then find
\b
G=d\psi^4(F),
\e
for some arbitrary constant $d$, where we must have $d\psi(F)\neq0$.
The Sundman transformation (\ref{sund}) thus yields a mapping from (\ref{p4W}), i.e.,
\b
W_{XX}=\frac{1}{2}\frac{W_X^2}{W}+\frac{\beta}{d^2}\frac{1}{W}+4dF\psi^6(F)W^2
+\frac{3}{2}d^2\psi^8(F)W^3,
\label{p4Wa}
\e
to $P_{IV}$. We note that equation (\ref{p4Wa}) depends on four arbitrary constants: 
$c$, $d$, and two arbitrary constants $a$ and $b$ which appear via  the general 
solution of (\ref{p4psieqn}). This may be expressed in terms of Weber-Hermite or
parabolic cylinder functions. It is interesting that such functions, relevant to
the study of $P_{IV}$, occur here.

\setcounter{equation}{0}

\section{Alternative third order autonomous equations and Sundman transformations}

We consider in this section an alternative approach to deriving third order 
autonomous equations related to the four Painlev\'e equations $P_I$-$P_{IV}$.
Here the idea is to directly eliminate the 
independent variable $z$ between the equation and its derivative. This we do
by expressing both of these equations as polynomials in $z$, and is a purely 
algebraic process. The dependent variable of the third order autonomous 
equation thus obtained is then precisely that of the Painlev\'e  equation 
itself. We may then apply the Sundman transformation (\ref{sund}) to this 
third order autonomous equation, and proceed as in the previous section to 
identify the transformed equation with a second order nonautonomous equation.
Here we will undertake this process only for $P_I$ and $P_{II}$, as the
equations resulting from the application of the Sundman transformation 
(\ref{sund}) in the $P_{III}$ and $P_{IV}$ cases have many terms, with their
coefficients, expressed as previously in terms of $F$ and $G$, being very long.
It is still the case, however, that we regard the derivation of equivalent
autonomous equations, here in the same dependent variable, as of interest, and 
so give the corresponding results for $P_{III}$ and $P_{IV}$.

Let us give a general formulation of this elimination process, for second order
nonautonomous equations expressible as being linear in $z$ (e.g., $P_I$, $P_{II}$ 
and $P_{III}$), or as being quadratic in $z$ (e.g., $P_{IV}$). This then yields
equivalent autonomous equations. We may write the first class of equation in the 
form
\b
z+b=0,
\label{lin}
\e
where $b$ is a function of $y$, $y_z$ and $y_{zz}$. It trivially follows that 
such an equation is equivalent to the third order autonomous equation
\b
1+b_z=0.
\label{dlin}
\e
Clearly, the autonomous equation follows from the previous equation by derivation.
The equivalence of these two equations is due to the fact that any constant of
integration $z_0$ in the first integral
\b
z-z_0+b=0
\e
of (\ref{dlin}) can be removed by a shift on $z$: thus we see also that equation 
(\ref{dlin}) implies equation (\ref{lin}).

The process for equations quadratic in $z$ is similar. We assume the equation
to be in the form
\b
z^2+bz+c=0,
\label{quad}
\e
where $b$ and $c$ are functions of $y$, $y_z$ and $y_{zz}$. Derivation of this 
equation gives
\b
2z+b_z z +b+c_z=0,
\e
and so
\b
z=-\frac{b+c_z}{2+b_z},
\e
and substitution in (\ref{quad}) then yields the third order autonomous equation
\b
\left(\frac{b+c_z}{2+b_z}\right)^2-b\left(\frac{b+c_z}{2+b_z}\right)+c=0.
\label{dquad}
\e
The equivalence of (\ref{quad}) and (\ref{dquad}), i.e., that it is also the case
that the latter implies the former, follows similarly to the previous case of 
equations linear in $z$: any constant of integration $z_0$ in a first integral of
(\ref{dquad}) defined by
\b
(z-z_0)^2+b(z-z_0)+c=0
\e
can always be removed by a shift on $z$.

\subsection{The first Painlev\'e equation}

The first Painlev\'e equation,
\b
y_{zz}=6y^2+z,
\e
is trivially transformed by derivation onto the third order equation
\b
y_{zzz}-12yy_z-1=0.\label{derP1}
\e
The application of the Sundman transformation (\ref{sund}), now written
\b
y(z)=F(X),\qquad  dz=G(X)dt, \qquad {\rm where} \qquad GF_X\neq0,
\label{sunda}
\e
to equation (\ref{derP1}) yields the following equation for the new 
variable $X(t)$:
\b
X_{ttt}+A(X)X_{t}X_{tt}+B(X)X_t^3+C(X)X_t+D(X)=0,\label{p1xt}
\e
where
\ba
A(X) &=& 3\frac{F_{XX}}{F_X}-4\frac{G_X}{G},\\
B(X) &=&\frac{F_{XXX}}{F_X}-3\frac{G_XF_{XX}}{GF_X}-\frac{G_{XX}}{G}+3\frac{G_X^2}{G^2},\\
C(X) &=&-12G^2F ,\\
D(X)&=&-\frac{G^3}{F_X},
\ea
and where we must have both $C(X)$ and $D(X)$ different from zero.
As in the previous section, we write equation (\ref{p1xt}) as the system
\ba
X_t&=&Y,\\
Y_t&=&Z,\\
Z_t&=&-A(X)YZ-B(X)Y^3-C(X)Y-D(X).
\ea
The equation satisfied by $Y=Y(X)$ is then found to be
\b
YY_{XX}=-Y_X^2-A(X)YY_X-B(X)Y^2-C(X)-\frac{D(X)}{Y},
\e
and the corresponding equation satisfied by $W(X)=1/Y(X)$ is
\b
W_{XX}=3\frac{W_X^2}{W}-A(X)W_X+B(X)W+C(X)W^3+D(X)W^4.
\e

Let us now consider the case where we require $A(X)=B(X)=0$. It follows
that $F(X)$ satisfies
\b
\left(\frac{F_{XX}}{F_X}\right)_X-\frac{1}{2}\left(\frac{F_{XX}}{F_X}\right)^2=0,
\e
and so
\b
F(X)=\frac{aX+b}{cX+d},
\e
where $a$, $b$, $c$ and $d$ are arbitrary constants. Since we must have at least
one of $c$ and $d$ different from zero, the above expression contains just three
arbitrary constants (the quotients obtained by dividing by nonzero $c$ or $d$).
We note that the condition $F_X\neq0$ corresponds to a requirement $ad-bc\neq0$.
We then obtain $G^4=eF_X^3$, where $e\neq0$ is a fourth arbitrary constant. The 
Sundman transformation (\ref{sunda}) then provides a mapping from
\b
W_{XX}=3\frac{W_X^2}{W}+C(X)W^3+D(X)W^4,
\e
where $C(X) =-12G^2F$ and $\displaystyle D(X)=-\frac{G^3}{F_X}$, with $F(X)$ and
$G(X)$ as above, to $P_I$.

\subsection{The second Painlev\'e equation}

Eliminating $z$ between the second Painlev\'e equation,
\b
y_{zz}=2y^3+zy+\alpha,
\e
and its derivative yields the equation
\b
yy_{zzz}-4y^3y_z-y_zy_{zz}-y^2+\alpha y_z=0.\label{derP2}
\e
Applying the Sundman transformation (\ref{sunda}) to equation (\ref{derP2}), 
we obtain
\b
X_{ttt}+A(X)X_{t}X_{tt}+B(X)X_t^3+C(X)X_t+D(X)=0,\label{p2xt}
\e
where
\ba
A(X) &=& 3\frac{F_{XX}}{F_X}-4\frac{G_X}{G}-\frac{F_X}{F},\\
B(X) &=&\frac{F_{XXX}}{F_X}-3\frac{G_XF_{XX}}{GF_X}-\frac{G_{XX}}{G}+3\frac{G_X^2}{G^2}-\frac{F_{XX}}{F}+\frac{F_XG_X}{FG},\\
C(X) &=&\frac{G^2}{F}(\alpha-4F^3),\\
D(X)&=&-\frac{G^3 F}{F_X},
\ea
and where we must have both $C(X)$ and $D(X)$ different from zero. Since equation
(\ref{p2xt}) has the same form as equation (\ref{p1xt}), albeit with different
coefficients, it follows that the form of the corresponding equation in $W(X)$
must also be as derived for the first Painlev\'e equation, i.e.,
\b
W_{XX}=3\frac{W_X^2}{W}-A(X)W_X+B(X)W+C(X)W^3+D(X)W^4.
\label{p25W}
\e

If for this last equation we impose a requirement that $A(X)=B(X)=0$, we find that
$F$ must satisfy
\b
\left(\frac{F_{XX}}{F_X}\right)_X-\frac{1}{2}\left(\frac{F_{XX}}{F_X}\right)^2
=\frac{3}{2}\frac{1}{F^2}F_X^2,
\label{p25Fsch}
\e
which is of the form (\ref{Fsch}) with $\displaystyle g(F)=\frac{3}{F^2}$. Thus, 
from Lemma 4.2 we obtain the general solution of (\ref{p25Fsch}) as
\b
X+c=\int\frac{dF}{\psi^2(F)},
\label{p25XcF}
\e
where $c$ is an arbitrary constant and $\psi(F)$ is the general solution of the
Euler equation
\b
\frac{d^2\psi}{dF^2}=\frac{3}{4}\frac{1}{F^2}\psi.
\label{p25psieqn}
\e
We then find
\b
G=d\left(\frac{\psi^6(F)}{F}\right)^\frac{1}{4},
\e
for some arbitrary constant $d$, where we must have $d\psi(F)\neq0$.
The Sundman transformation (\ref{sunda}) thus yields a mapping from (\ref{p25W}), i.e.,
\b
W_{XX}=3\frac{W_X^2}{W}+d^2\left(\frac{\psi^6(F)}{F}\right)^\frac{1}{2}
\frac{(\alpha-4F^3)}{F}W^3-d^3\left(\frac{\psi^6(F)}{F}\right)^\frac{3}{4}
\frac{F}{\psi^2(F)}W^4.
\label{p25Wa}
\e
to $P_{II}$. We note that equation (\ref{p25Wa}) depends on four arbitrary 
constants: $c$, $d$, and two arbitrary constants $a$ and $b$ which appear 
via the general solution $\psi(F)=aF^{-\frac{1}{2}}+bF^\frac{3}{2}$ of 
(\ref{p25psieqn}).

\subsection{The third and fourth Painlev\'e equations}

As indicated earlier, for $P_{III}$ and $P_{IV}$ we simply record the autonomous
equations obtained by elimination of $z$ between the equation and its derivative.
For $P_{III}$ we obtain
\ba
&&(y^2y_z-\alpha y^4-\beta y^2)y_{zzz}-2y^2y_{zz}^2+\left[yy_z^2+2(2\alpha y^2+\beta )yy_z+3y(\gamma y^4+\delta )\right]y_{zz}-(3\alpha y^2+\beta)y_z^3\nonumber\\
&&-(5\gamma y^4+\delta)y_z^2+
(\alpha\gamma y^6+3\gamma\beta y^4-3\alpha \delta y^2-\beta\delta)y_z
-\gamma^2y^8-2\gamma\delta y^4-\delta^2=0,\label{derP3}
\ea
and for $P_{IV}$ we obtain
\ba
&&y^4y_{zzz}^3+2\left[-2yy_{zz}+y_z^2+(y^4+2\beta )\right]y^2y_zy_{zzz}
+4y^2y_z^2y_{zz}^2-4\left[y_z^4+(3y^4+2\beta)y_z^2+4y^4y_z+2y^4\right]yy_{zz}
\nonumber\\
&&+y_z^6+2(3y^4+2\beta)y_z^4+8y^4y_z^3
+(-3y^8-16\alpha y^6+12\beta y^4+4y^4+4\beta^2)y_z^2\nonumber\\
&&+4y^4(2\beta-y^4-4\alpha y^2)(2y_z+1)=0.\label{derP4}
\ea
We believe these autonomous equations, equivalent to $P_{III}$ and $P_{IV}$,
to be of interest. We will return to a study of these equations, as well as 
of similarly-derived equations for other Painlev\'e equations, in later papers.

\setcounter{equation}{0}

\section{Conclusions and Discussion}

In this paper we have explored the application of Sundman transformations to the
Painlev\'e equations, in particular to the first four Painlev\'e equations
$P_I$-$P_{IV}$, with $P_V$ and $P_{VI}$ being left to a later paper. This requires
first of all the construction of an equivalent autonomous system, a process which
we have discussed in general in Section 2. This step we characterize as being
inverse to the use of Lie symmetries to obtain a reduction of order of an autonomous
system. In \cite{E07} this step was made, for the case of the first Painlev\'e 
equation, using a specific hodograph transformation. The second step is to apply a
Sundman transformation to the autonomous system, and to identify the resulting
equation with a corresponding nonautonomous equation. A mapping between solutions
of this nonautonomous equation and the original Painlev\'e equation may then be given
explicitly, which we have done in equation (\ref{smap}). This explicit mapping is 
not to be found in \cite{E07}. Interesting special cases of the nonautonomous equation 
found by Sundman transformation have also been considered, just as the special case 
of an Emden-Fowler equation was considered in \cite{E07}. Thus, for $P_{II}$, we obtain
a nonautonomous equation with coefficients expressed via Airy functions, and for
$P_{IV}$ a nonautonomous equation with coefficients expressed via Weber-Hermite
functions; it is interesting that these functions, well-known to respectively play
an important role in the study of these equations, should also appear here. Finally,
we have discussed an alternative, more direct, autonomisation process, and have used
this to again apply Sundman transformations to $P_I$ and $P_{II}$.

Our results lead to a number of possible future directions and interesting open 
problems. One possible future direction is to consider other means of autonomising
a given nonautonomous equation. In addition to that discussed in section 5, the
further application of Theorem 2.7, as in Section 3.5 for $P_{III}$ and $P_{IV}$,
might also be considered, although in general this will lead to an increase in order 
by more than one. Other approaches to autonomisation include, for example, that 
employed by Noumi in \cite{N04}. A second future direction is the further study of the
nonautonomous equations obtained by application of a Sundman transformation. A third
direction to explore is the use of a more general Sundman transformation, as noted 
in \cite{E07} (see Remark 4.1). 

We also expect to return to a study of the autonomous equations obtained here in their
own right. One starting point would be the application, in the case of polynomial
autonomous systems, of algebraic techniques such as those used in \cite{CF06,Bou1,Bou2}.
A second topic for future discussion in the construction of equivalent autonomous
equations is the preservation or otherwise of the Painlev\'e property. We note, for
example, that the technique discused in Section 5 gives rise to autonomous equations
in the same dependent variable, as does the approach of Noumi. Such equations could
then be obtained using a Painlev\'e classification of third order autonomous equations.
This is not true, however, of the equations obtained from the autonomisation process
presented in Section 2, as this will not, in general, preserve the Painlev\'e property.

This then leads to the question of how, in a study of third order autonomous systems,
or scalar equations, we might isolate such equivalent systems or equations as discussed 
in Section 3. What properties mark out such systems and equations as special, thus
allowing us to determine that, of all the equations in some general class, they are of
particular interest? The equivalence to Painlev\'e equations guarantees, for example,
that they are integrable. We may thus give Lax pairs for such equivalent autonomous
systems. For example, the system
\ba
v_{0,x} & = & g_0(v_0,v_1,v_2), \\
v_{1,x} & = & g_0(v_0,v_1,v_2)v_2, \\
v_{2,x} & = & g_0(v_0,v_1,v_2)(6v_1^2+v_0),
\ea
which corresponds to the general system (\ref{paut1})---(\ref{paut3}) in the $P_I$ case, 
i.e., with $F(v_0,v_1,v_2)=6v_1^2+v_0$ (and any $g_0(v_0,v_1,v_2)\not\equiv 0$), has 
the Lax pair
\b
\Psi_x=g_0
\left(\begin{array}{cc} 0 & 1 \\ \lambda+2v_1 & 0\end{array}\right)\Psi,\qquad{}
\Psi_\lambda=
\left(\begin{array}{cc} v_2 & 2\lambda-2v_1 \\ 2\lambda^2+2\lambda v_1 +2v_1^2+v_0 &
-v_2\end{array}\right)\Psi.
\e
The compatibility condition $g_0F_\lambda-G_x+g_0[F,G]=0$ of this Lax pair then gives
precisely the above autonomous system. We note that this Lax pair has (as we might
expect) a form similar to those of Painlev\'e equations, rather than the form usually
considered for autonomous systems (similar Lax pairs may also be given for autonomous
equations equivalent to the other Painlev\'e equations). But how would we identify 
any given system of the above form as integrable without explicitly construcing its
Lax pair? What criteria might we use? We might be able to deduce the transformation to $P_I$, but given that any choice of $g_0(v_0,v_1,v_2)$ may be made, this will not
necessarily be straightforward. Furthermore, here we are considering systems and
equations equivalent to the Painlev\'e equations themselves, but the same questions 
may be asked of systems equivalent to higher order Painlev\'e equations, the Sundman
transformations of which is itself a topic worthy of future research. In some sense,
this brings us back to the ARS conjecture \cite{ARS80}, that all integrable partial
differential equations will have reductions to ordinary differential equations with
the Painlev\'e property, perhaps after a change of variable. Whilst allowing a change 
of variable may be necessary in order to cater for examples such as the Harry Dym 
equation, it remains the case that, when we do not know to expect such a change of
variable (when the partial differential equation is known to be integrable, like Harry
Dym), then when a reduction does not have the Painlev\'e property it remains unclear 
how to proceed: do we seek to prove nonintegrability, or do we instead seek a change 
of variable to an equation with the Painlev\'e property, and if so what class of
change of variable?

\section*{Funding}

The authors thank the Universidad Rey Juan Carlos for funding through the project
2025/SOLCON-160677. PRG and AP also thank the Universidad Rey Juan Carlos for 
funding as members of the Grupo de Investigaci\'on de Alto Rendimiento DELFO, and
AI as member of the Grupo de Investigaci\'on de Alto Rendimiento ALCRYPT and
collaborator of the Grupo de Investigaci\'on Emergente COMESTOP.

\section*{Competing interests}

The authors have no competing interests to declare that are relevant to the content
of this article.


\begin{thebibliography}{99}
	
\bibitem{Sund13} Sundman, K. F.,  M\'emoire sur le probl\`eme des trois corps. {\sl Acta Math.}  36, 105-179 (1913).
	
\bibitem{DMS94} Duarte, D. G. S., Moreira, I. C., Santos, F. C., Linearization under non-point transformations. {\sl J. Phys. A: Math. Gen.}  27, L739-L743 (1994).
	
\bibitem{KS14} Kudryashov, N. A., Sinelshchikov, D. I.,  Analytical solutions of the Rayleigh equation for empty and gas filled bubble. {\sl J. Phys. A: Math and Theor.}  47, 405202 (2014).

\bibitem{KS15} Kudryashov, N. A., Sinelshchikov, D. I.,  Analytical solutions for problems of bubble dynamics. {\sl Phys. Lett. A}  379, 798-802 (2015).

\bibitem{MR16} Mancas, S.C., Rosu, H.C., Cavitation of spherical bubbles: closed-form, parametric and numerical solutions. {\sl Physics of Fluids}, 28 (2), 022009 (2016).

\bibitem{GP21} Gordoa, P. R., Pickering, A., Ultrasonic Waves in Bubbly Liquids: An Analytic Approach. {\sl Mathematics} 9, 1309 (2021).

\bibitem{GPPT25}  Gordoa, P. R., Pickering, A., Puertas-Centeno, D., Toranzo, E. V., Generalized and new solutions of the NRT nonlinear Schrödinger equation. {\sl Physica D: Nonlinear Phenomena} 472, 134515 (2025).

\bibitem{GPPT26}  Gordoa, P. R., Pickering, A., Puertas-Centeno, D., Toranzo, E. V.,
Sundman-like transformations and the NRT nonlinear Schrödinger equation.
{\sl Mathematical Methods in the Applied Sciences} 49, 14791-14810 (2026).

\bibitem{E07} Euler, M., Euler, N., Str\"omberg, A., \AA str\"om, E., Transformation
between a generalized Emden-Fowler equation and the first Painlev\'e transcendent.
{\sl Mathematical Methods in the Applied Sciences} 30, 2121-2124 (2007).

\bibitem{CF06} Carr\'a Ferro, G., Generalized differential resultant systems of 
algebraic ODEs and differential elimination theory. ``Trends in Mathematics: 
Differential Equations with Symbolic computation,'' eds. Wang, D., Zheng, Z.,
327-341. Birkh\"auser Verlag, Basel (2005).

\bibitem{RR10} Retakh, V., Rubtsov, V., Noncommutative Toda chain, Hankel
quasideterminants and Painlev\'e II equation. {\sl J. Phys. A: Math. Theor.} 43,
505204 (2010).

\bibitem{W84} Weiss, J., B\"acklund transformation and the H\'enon-Heiles system.
{\sl Phys. Lett. A} 105, 387-389 (1984).

\bibitem{gu1} Guha, P., Khanra, B., Choudhury, A. G., On generalized Sundman
transformation method, first integrals, symmetries and solutions of equations
of Painlev\'e-Gambier type. {\sl Nonlinear Analysis} 72, 3247-3257 (2010).

\bibitem{gu2} Guha, P., Choudhury, A. G., Khanra, B., On solutions of third and
fourth-order time dependent Riccati equations and the generalized Chazy system.
{\sl Commun. Nonlinear Sci. Numer. Simulat.} 17, 4053-4063 (2012).

\bibitem{PZ02} Polyanin, A. D., Zaitsev, V. F., ``Handbook of Exact Solutions for Ordinary Differential Equations,'' 2nd edition. CRC Press, Boca Raton (2002).

\bibitem{LR15} Llibre, J., Rodrigues, A., A non-autonomous kind of Duffing equation.
{\sl Applied Mathematics and Computation} 251, 669-674 (2015).

\bibitem{GLS02} Gromak, V. I., Laine, I., Shimomura, S., ``Painlev\'e Differential
Equations in the Complex Domain.'' Walter de Gruyter, Berlin (2002).

\bibitem{N04} Noumi, N., ``Painlev\'e Equations through Symmetry.'' Translations of
Mathematical Monographs, vol. 223, American Mathematical Society, Providence, Rhode
Island (2004).

\bibitem{Bou1} Boulier, F., Differential elimination and biological modelling.
''Gr\"obner Bases in Symbolic Analysis,'' eds. Rosenkranz, M., Wang, D., 109-137.
 Walter de Gruyter, Berlin (2007).

\bibitem{Bou2} Boulier, F., Lemaire, F., Differential algebra and QSSA methods in
biochemistry. {\sl IFAC Proceedings Volumes} 15th IFAC Symposium on System 
Identification, 33-38 (2009).
	
\bibitem{ARS80} Ablowitz, M. J., Ramani, A., Segur, H., A connection between nonlinear evolution equations and ordinary differential equations of P-type. I. {\sl J. Math. 
Phys.} 21, 715-721 (1980).
	
\end{thebibliography}
\end{document}